# Sulvanite Compounds $Cu_3TMS_4$ ($TM$= V, Nb and Ta): Elastic, Electronic, Optical and Thermal Properties using First-principles Method

**M. A. Ali [1*], N. Jahan [1], and A. K. M. A. Islam [2]**

[1]Department of Physics, Chittagong University of Engineering and Technology, Chittagong-4349, Bangladesh

[2]International Islamic University Chittagong, 154/A College Road, Chittagong, Bangladesh

## Abstract

We present a systematic first-principles study of the structural, elastic, electronic, optical and thermodynamics properties of the sulvanite compounds $Cu_3TMS_4$ ($TM$ = V, Nb and Ta). The structural, elastic and electronic properties are in fact revisited using a different calculation code than that used by other workers and the results are compared. The band gaps are found to be 1.041, 1.667 and 1.815 eV for $Cu_3VS_4$, $Cu_3NbS_4$ and $Cu_3TaS_4$, respectively which are comparable to other available calculated results. The optical properties such as dielectric function, refractive index, photoconductivity, absorption coefficients, reflectivity and loss function have been calculated for the first time. The calculated results are compared with the limited measured data on energy dependent refractive index and reflectivity coefficient available only for $Cu_3TaS_4$. All the materials are dielectric, transparent in the visible range. The values of plasma frequencies are found to be 15.36, 15.58 and 15.64 eV for $Cu_3VS_4$, $Cu_3NbS_4$ and $Cu_3TaS_4$, respectively. Furthermore, following the quasi-harmonic Debye model, the temperature effect on the bulk modulus, heat capacity, and Debye temperature is calculated reflecting the anharmonic phonon effects and these are compared with both experimental and other theoretical data where available.

## 1. Introduction

The ternary compounds $Cu_3TMS_4$ ($TM$ =V, Nb, Ta) have been reported to be interesting materials which exhibit promising opto-electronic properties with relatively large optical band gap, p-type conductivity, photoemission in the visible range [1-2]. Despite the prospect only a few experimental and theoretical works have been reported in the literature [3-8]. Thermodynamic and experimental investigation on the mixed conductivity in Cu$_3$VS$_4$ [3], infrared-reflectivity and Raman study of the vibronic properties of Cu$_3$*TM*S$_4$ (*TM* = V, Nb, Ta) [4], experimental analysis of the optical and transport properties of Cu$_3$TaS$_4$ thin film [1-2] have been reported. The elastic, electronic and bonding properties are studied theoretically [6-8].

---

[*] *Corresponding author*: ashrafphy31@gmail.com

We know the optical properties are directly related to the electronic band structure. Moreover, the thermodynamic properties are the basis of solid-state science and industrial applications since they can extend our knowledge on the specific behavior of materials under high pressure and high temperature environments [9]. To our knowledge, theoretical studies on the optical and thermodynamic properties of these materials have not been done yet. So, we are encouraged to investigate the detailed optical and thermodynamics properties of Cu$_3$*TM*S$_4$ (*TM* =V, Nb, Ta). In order to take the advantages of these compounds for eventual technological applications, an overall theoretical investigation is necessary. Therefore, in this paper, a systematic investigation of the optical and thermodynamic properties of these materials using a first-principles method has been performed within the generalized gradient approximation (GGA) for the exchange-correlation potential. Further a revisit on the elastic and electronic properties of the compounds will also be carried out using a different calculation code than hitherto been done by other workers via first principles density functional theory using the projector augmented wave (PAW) method and the GGA as implemented in the VASP code [6-8].

## 2. Computational Method

The calculations are carried out utilizing the GGA method with the Perdew, Burke and Ernzerhoff (GGA-PBE) [10] form to describe the exchange and correlation potential based on the density functional theory (DFT) [11] as implemented in the CASTEP code [12]. A plane-wave cutoff energy of 500 eV was used and the first brillouin zone integration was performed using Monkhorst–Pack scheme k-point sampling [13]. We had set the energy calculation to ultrafine quality with k-point mesh of $6 \times 6 \times 6$ for the crystal structure optimization. This basis sets makes the tolerance for self-consistent field, energy, maximum force, maximum displacement and maximum stress to be $5.0\times10^{-7}$ eV/atom, $5.0\times10^{-6}$ eV/atom, 0.01eV/Å, $5.0\times10^{-4}$ Å and 0.02GPa, respectively.

The *E-V* data necessary as input in the Gibbs program [14] which is based on the quasi-harmonic Debye model are calculated from the third-order Birch-Murnaghan equation of state (EOS) [15] using the zero temperature and zero pressure equilibrium values of $E_0$, $V_0$, and $B_0$, obtained through the DFT method, and allows us to calculate the bulk modulus, Debye temperature, heat capacities at any temperature and pressures. A detailed description of the quasi-harmonic Debye model can be found in Ref. [14].

## 3. Results and Discussion

### 3.1. *Structural and Elastic properties*

The compounds $Cu_3TMS_4$ ($TM$ =V, Nb, Ta) belong to the cubic systems with space group $P\bar{4}3m$ (215) consisting of Cu, TM, and S atoms which are positioned at: $3d$ (0.5, 0, 0), $1a$ (0, 0, 0) and $4e$ ($u, u, u$) sites of Wyckoff coordinates, respectively. The total energy is minimized by the geometry optimization and the optimized values of structural parameters of $Cu_3TMS_4$ are shown in Table 1 along with other theoretical results. The obtained lattice parameters are in good agreement with the results obtained by Osorio-Guillén *et al.* [8].

Table 1. The optimized structural parameters ($a, u$), independent elastic constants $C_{ij}$, bulk modulus $B$, shear modulus $G$, Young's modulus $Y$, Poisson ratio $v$, Zener's anisotropy index $A$, Pugh ratio $G/B$ of $Cu_3TMS_4$ ($TM$ =V, Nb, Ta).

| Parameters | $Cu_3VS_4$ | $Cu_3NbS_4$ | $Cu_3TaS_4$ |
|---|---|---|---|
| $a$ (Å) | 5.4213[a] | 5.5292[a] | 5.5622[a] |
| | 5.4374[b] | 5.5492[b] | 5.5588[b] |
| | 5.393[c] | 5.5001[d] | 5.5145[e] |
| $u$ | 0.2362[a], | 0.2420[a], | 0.2436[a], |
| | 0.2356[b] | 0.2421[b] | 0.2409[b] |
| | 0.2376[c] | 0.2426[d] | 0.2475[e] |
| $C_{11}$ (GPa) | 92.4[a], 92.1[f] | 97.8[a], 91.6[f] | 96.2[a], 89.0[f] |
| $C_{12}$ (GPa) | 16.0[a], 17.3[f] | 15.6[a], 12.0[f] | 11.0[a], 12.6[f] |
| $C_{44}$ (GPa) | 26.2[a], 20.4[f] | 22.3[a], 17.9[f] | 23.6[a], 17.5[f] |
| $B$ (GPa) | 41.4[a], 42.2[f] | 43.0[a], 38.5[f] | 39.4[a], 38.1[f] |
| $G$ (GPa) | 30.5[a], 27.08[f] | 28.5[a], 24.54[f] | 30.0[a], 24.06[f] |
| $Y$ (GPa) | 73.40[a], 66.92[f] | 70.00[a], 60.72[f] | 71.70[a], 59.62[f] |
| $v$ | 0.20[a] | 0.22[a] | 0.19[a] |
| $G/B$ | 0.73[a], 0.64[f] | 0.66[a], 0.63[f] | 0.76[a], 0.63[f] |
| $A$ | 0.68[a], 0.50[f] | 0.54[a], 0.40[f] | 0.54[a], 0.50[f] |

a: Present study, b: [8], c:[17], d:[18], e:[5], f:[7]

The elastic parameters (independent elastic constants $C_{ij}$, bulk moduli $B$, shear moduli $G$, Young's moduli $Y$, the Poisson ratio $v$, Zener's anisotropy index, $A$ and Pough ratio) of $Cu_3TMS_4$ ($TM$ =V, Nb, Ta) have been calculated using a different calculation code (CASTEP code). The computed results are presented in Table 1 along with theoretical data obtained by Espinosa-García [7] for comparison. The obtained elastic constants satisfy the general criteria [16] for mechanical stability: $(C_{11}-C_{12}) > 0$; $(C_{11} + 2C_{12}) > 0$; $C_{44} > 0$. These conditions also lead to a restriction on the value of the bulk modulus $B$ as $C_{12} < B < C_{11}$, again satisfied by the present data. The table further shows that our optimized structural data are in good agreement with the available results obtained by other workers [5,7-8,17-18].

The theoretical polycrystalline elastic moduli for $Cu_3TMS_4$ (*TM* =V, Nb, Ta) may be computed from the set of independent elastic constants. Hill's [19] proved that the Voigt and Reuss equations represent upper and lower limits of the true polycrystalline constants. The bulk modulus $B$ and shear modulus $G$ are obtained from elastic constants according to the Voigt-Reuss-Hill (VRH) average scheme [19]. The subscript V denotes the Voigt bound, R denotes the Reuss bound. The arithmetic average of Voigt and Reuss bounds is termed as the Voigt-Reuss-Hill approximations [20] $B_H \equiv B = ½(B_R + B_V)$ and $G_H \equiv G = ½(G_R + G_V)$. To complete the elastic properties we have calculated the Young's modulus, $Y$ and Poisson's ratio, $v$ by using the expressions which can be found elsewhere [21-22]. Furthermore, the Zener's anisotropy index, $A$ is also calculated by using the expression, $A = 2C_{44}/(C_{11}-C_{12})$ [23].

The bulk modulus $B$ describes the resistance of a material to volume change. The values of bulk modulus follow the sequence is $B (Cu_3NbS_4) > B (Cu_3VS_4) > B (Cu_3TaS_4)$ which are in good agreement with those of Ref. [7]. The Young's modulus $Y$, denoting a measure of stiffness, shows that $Cu_3VS_4$ is stiffer than the other two phases. Frantsevich [24] gave the critical value of Poisson's ratio of a material is 0.33. For brittle materials such as ceramics, the Poisson's ratio is less than 0.33, otherwise, the material is ductile, and it could be seen that the examined compounds are brittle materials. Pugh [25] proposed a critical value for ductile-brittle transition. If $G/B > 0.5$, the material behaves as a brittle material, otherwise it is ductile. From Table 1, we find that all the compounds are brittle in agreement with the results reported in Ref. [7]. Further the compounds are anisotropic in nature as the degree of anisotropy can be inferred from Zener's anisotropy index shown in the table.

### 3.2. *Electronic properties*

As mentioned earlier a revisit on the electronic properties has been made in order to assess the validity of our results with the available results. The calculated electronic band structure of $Cu_3TMS_4$ (*TM* =V, Nb, Ta) along high-symmetry directions of the crystal Brillouin zone (BZ) are displayed in Fig. 1 (left panel). The Fermi level is chosen to be zero of the energy scale. From the figure, it is clear that all the three compounds are semiconductors with indirect bandgap and the calculated band gap values are 1.041, 1.667 and 1.815 eV for $Cu_3VS_4$, $Cu_3NbS_4$ and $Cu_3TaS_4$, respectively. These values are in good agreement with those found by Osorio-Guillén *et al.* [8]. The calculated values are smaller in comparison with the experimental values of 1.3 [4] and 2.7 [2] eV for $Cu_3VS_4$ and $Cu_3TaS_4$, respectively. The underestimation of the band gap is a well-known effect of the DFT calculation and can be corrected by several approaches. One way is to identify the Kohn–Sham eigenvalues with quasi-particle energies. This can be accounted for by a rigid shift of the conduction band upwards with respect to the valence band [26-27]. Now for all the three compounds the bands that are between -6.5 to -3 eV below the Fermi level are derived mainly from S-*p*-states. The maximum valence band is located at the R point (1/2, 1/2, 1/2) and Cu-*d*-states are dominant and the minimum conduction band is located at the X-point (1/2, 0, 0) and its character is dominated by *TM* (*TM* = V, Nb, Ta)-*d*-states. For the compounds under consideration we can see that when we go down in the column *V-B* of the periodic table, the band gap values are found to increase. As the VBM in these materials remains fixed with reference to the Fermi level and the CBM is shifted to higher energies, the consequences of these are the increase in band gap of the sulvanite compounds.

The corresponding total and partial density of states are shown in Fig. 1 (*right panel*). From the total and partial density of states (PDOS) of $Cu_3VS_4$, $Cu_3NbS_4$, and $Cu_3TaS_4$, more details of the electronic structure of these materials are observed. For all the compounds the band structures around Fermi level, $E_F$ is derived mainly from Cu-*d* and S-*p* orbitals. The lowest valence band is dominated by S-*s* states. Whereas the top of the valence band down to around -5.7 eV is divided into two subbands and upper subband is dominated by Cu-*d* states down to around -2.6 and other subband is dominated by S-*p* states when a small contribution of transition metal *TM d*-states is noticeable. The lower energy subband 'down valence band' shows two main peaks localized due to S-*p*-states and Cu-*d* states. There are also two main peaks at the up valence band ranging from -2.6 eV to Fermi level due to Cu-*d* states while contributions from *TM-d* and S-*p* states are negligible. *TM-d* states are the most dominant at lowest conduction bands (CBs).

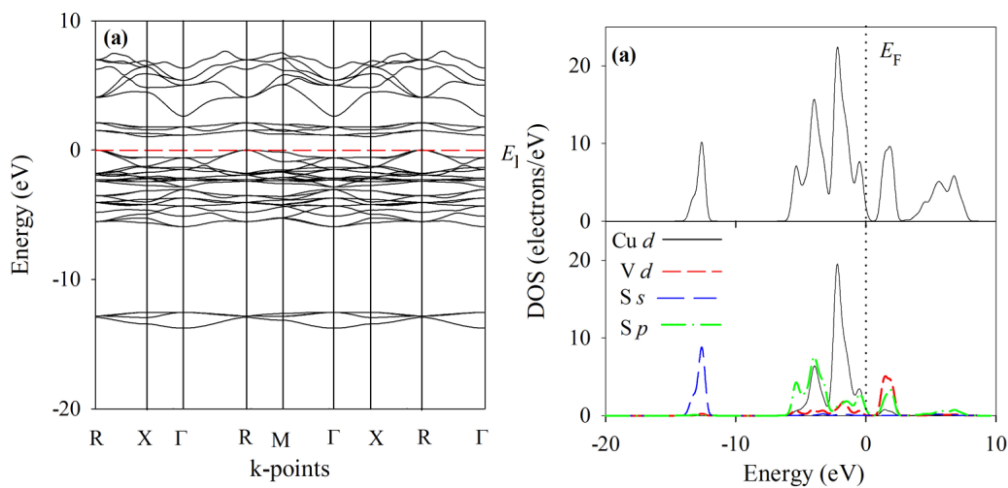

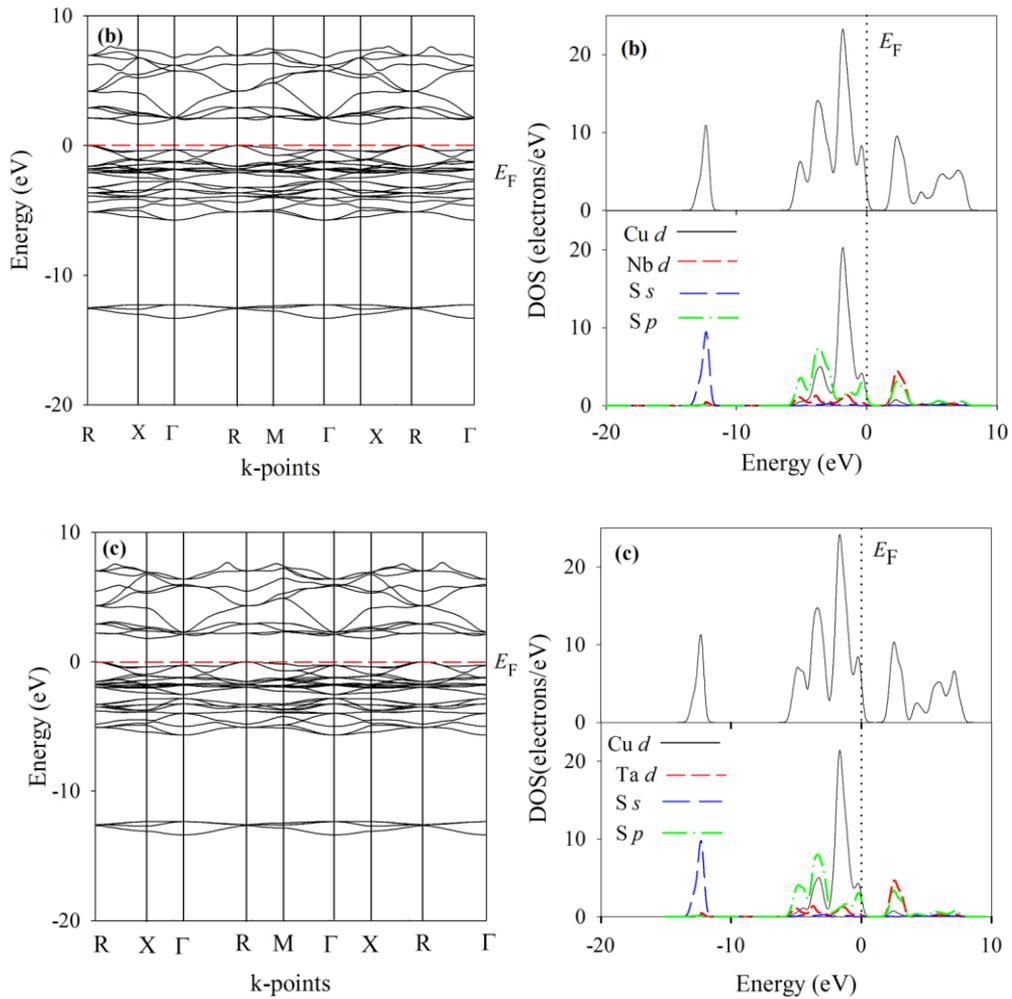

Fig. 1. The electronic band structure (left panel) and DOS of states (right panel) of (a) $Cu_3VS_4$, (b) $Cu_3NbS_4$ and (c) $Cu_3TaS_4$, respectively.

### 3.3. *Optical properties*

The optical properties of $Cu_3TMS_4$ (TM =V, Nb, Ta) are studied by the complex dielectric function, $\varepsilon(\omega) = \varepsilon_1(\omega) + i\,\varepsilon_2(\omega)$ which is one of the main optical characteristics of solids. The imaginary part $\varepsilon_2(\omega)$ is obtained in CASTEP numerically by a direct evaluation of the matrix elements between the occupied and unoccupied electronic states and calculated using the following equation:

$$\varepsilon_2(\omega) = \frac{2e^2\pi}{\Omega\varepsilon_0} \sum_{k,v,c} \left|\psi_k^c|\hat{u}.r|\psi_k^v\right|^2 \delta\left(E_k^c - E_k^v - E\right),$$

where, $\psi_k^c$ and $\psi_k^v$ are the conduction and valence band wave functions at $k$, respectively and $\omega$ is the light frequency, $\hat{u}$ is the vector defining the polarization of the incident electric field, $e$ is the electronic charge. The Kramers–Kronig transform of the imaginary part $\varepsilon_2(\omega)$ gives the real part $\varepsilon_1(\omega)$. From the real and imaginary parts of the dielectric function one can calculate other spectra, such as refractive index, absorption spectrum, loss-function, reflectivity and conductivity (real part) using the expressions given in Ref. [26].

The calculated optical properties of the $Cu_3TMS_4$ (*TM* =V, Nb, Ta) compounds from the polarization vectors (100) are shown in Figs. 2 - 4 for energy range up to 20 eV. To the best of our knowledge there are no theoretical optical properties available for the compounds in literature, excepting some limited measurements of energy dependent refractive index and reflectivity for $Cu_3TaS_4$ at very low energy [1-2].

The primary quantity that characterizes the electronic structure of any crystalline material is the probability of photon absorption, which is directly related to the imaginary part of the optical dielectric function $\varepsilon(\omega)$. Fig. 2 (left panel), displays the calculated real and imaginary parts of the dielectric function for $Cu_3VS_4$, $Cu_3NbS_4$ and $Cu_3TaS_4$ sulvanites. It is seen that the calculated curves for the linear optical components $\varepsilon_1(\omega)$ and $\varepsilon_2(\omega)$ for $Cu_3TMS_4$ (*TM* =V, Nb, Ta) all seem to be similar but not identical. There are some differences in the positions of the two main peaks which are due to the optical transitions. The curves for $\varepsilon_2(\omega)$ show that the first peaks occur at about 2.80, 3.36 and 3.36 eV for $Cu_3VS_4$, $Cu_3NbS_4$, and $Cu_3TaS_4$, respectively. These peaks are due to the direct optical transitions between the highest valence band and the lowest conduction band at the $\Gamma$-point *i.e.,* electron transition from Cu-*d* states to *TM*-*d* states. The second peaks at 5.77, 6.22 and 6.22 eV for $Cu_3TMS_4$ (*TM* =V, Nb, Ta) may leads due to the electron transition from the hybridized orbitals of Cu-*d*, S-*p* and TM-*d* to the second conduction band. The calculated static dielectric constants $\varepsilon_1(0)$ are found to have the values 8.56, 7.32, and 7.10 for $Cu_3TMS_4$ (*TM* =V, Nb, Ta), respectively. Therefore, it is reasonable to infer that these are dielectric materials.

The knowledge of the refractive index of an optical material is important for its use in optical devices such as photonic crystals, waveguides etc. The refractive index *n* and the extinction coefficient *k* are displayed in Fig. 2 (*right panel*). The static refractive index *n*(0) is found to have the values 2.93, 2.71 and 2.66 for $Cu_3VS_4$, $Cu_3NbS_4$, and $Cu_3TaS_4$, respectively. The variation of refractive index with incident photon energy is similar to $\varepsilon_1(\omega)$, while the extinction coefficient followed $\varepsilon_2(\omega)$ as can be seen in Fig. 2. The wavelength dependent refractive index of $Cu_3TaS_4$ thin films has been measured by Newhouse *et al.* [2] which is plotted along with our result (Fig. 2(c), *right panel*) for comparison. It can be seen that our calculated data roughly follows the available measured data at low energy.

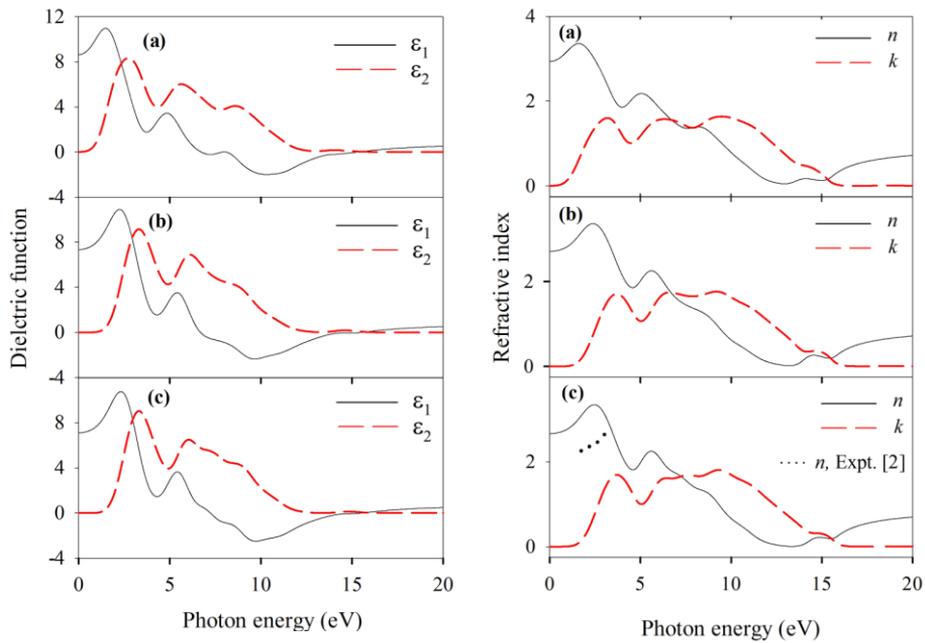

Fig. 2. The dielectric function (*left panel*) and refractive index (*right panel*) of (a) $Cu_3VS_4$, (b) $Cu_3NbS_4$ and (c) $Cu_3TaS_4$, respectively.

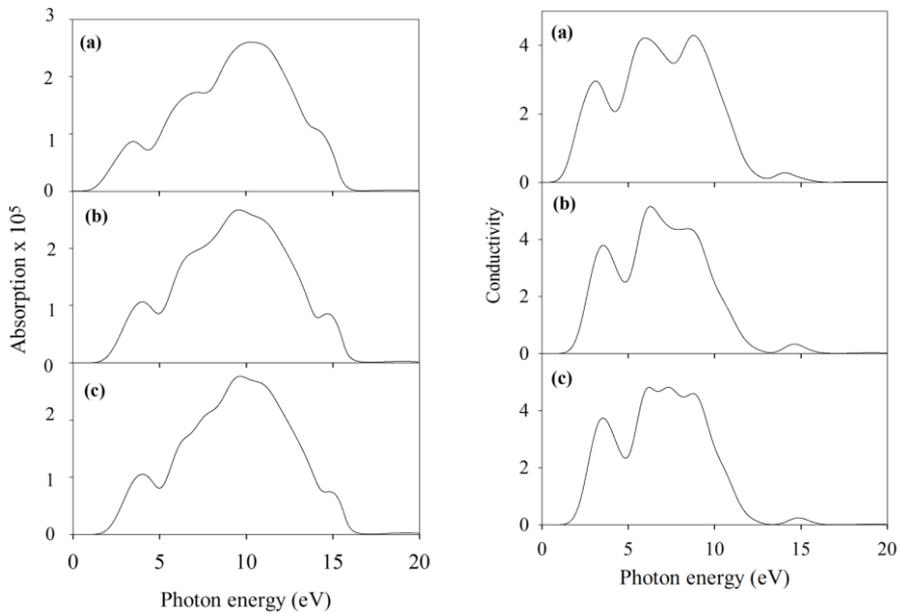

Fig. 3. The absorption coefficient (left panel) and conductivity (right panel) of (a) $Cu_3VS_4$, (b) $Cu_3NbS_4$ and (c) $Cu_3TaS_4$, respectively.

The absorption coefficient, displayed in Fig. 3 (left panel), shows that the structure for each of the studied compound has no absorption band at low energy range due to the semiconducting nature of the compounds. The materials have a sharp edge in their absorption coefficient, since light which has energy below the band gap does not have sufficient energy to raise an electron across the band gap. Fig. 3 (right panel) also shows that the photoconductivity starts with a zero value at low energy region due to the semiconducting nature and initiates photoconductivity due to the absorption of photon of sufficient energy resulting in the increase in the number of free carriers.

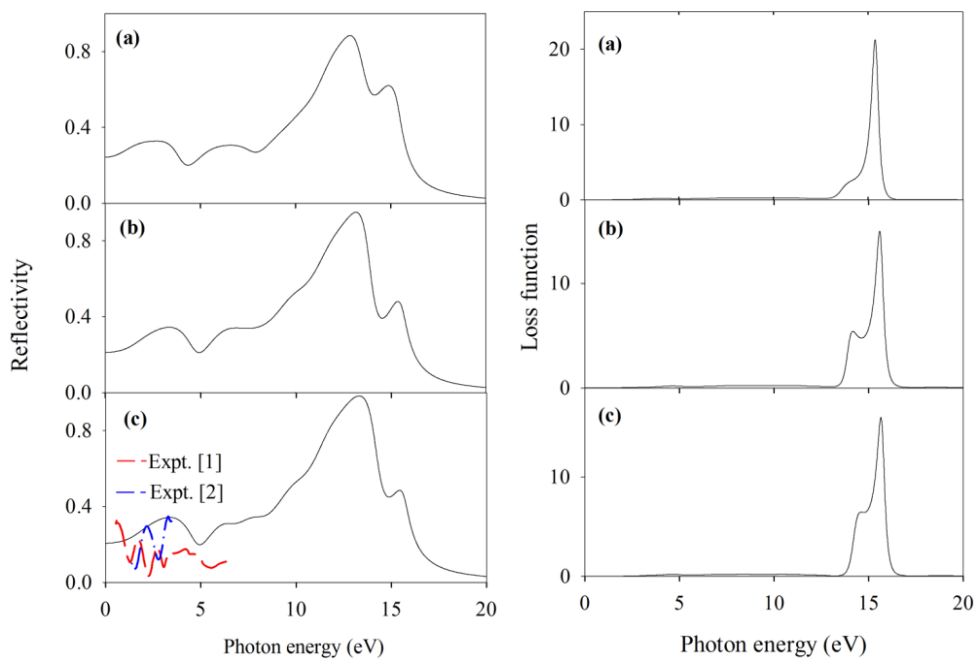

Fig. 4. The reflectivity (left panel) and loss function (right panel) of (a) $Cu_3VS_4$, (b) $Cu_3NbS_4$ and (c) $Cu_3TaS_4$, respectively. Two sets of the experimental data are shown in the left panel in Fig. 4(c) (see text).

Fig. 4 (left panel) shows the reflectivity spectra of $Cu_3TMS_4$ (*TM* = V, Nb, and Ta) as a function of photon energy. The small value of reflectance ~ 0.20-0.40 in the low energy region indicates that $Cu_3TMS_4$ (*TM* = V, Nb, and Ta) materials are mostly transmitting in this region. The reflectivity spectrum starts to rise sharply at ~ 7.0 eV reaches its maximum at 12.9 - 13.3 eV for the three compounds. Experimentally photoluminescence data on the spectral dependent of the reflectivity coefficient for 200 nm $Cu_3TaS_4$ film on a-$SiO_2$ at room temperature is available and is shown in Fig. 4(c) (left panel) [1-2]. The two experimental values differ from each other but are roughly of the order of the theoretical values. All these measured data originate from samples of thin films with $SiO_2$ substrates and these are only for very low energy [1-2]. So a meaningful comparison is hardly possible at this time.

Fig. 4 (right panel) shows the energy-loss function $L(\omega)$, which describes the energy loss of a fast electron traversing in the material. Its peaks represent the characteristics associated with the plasma resonance and the corresponding frequency is defined as the plasma frequency. The values of plasma frequencies are 15.36, 15.58, and 15.64 for $Cu_3VS_4$, $Cu_3NbS_4$, and $Cu_3TaS_4$, respectively. When the frequency of the light is higher than the plasma frequencies, the materials becomes transparent. The materials exhibit dielectric $[\varepsilon_1(\omega) > 0]$ and metallic $[\varepsilon_1(\omega) < 0]$ behaviors above and below the plasma frequency, respectively. Moreover, the peak in loss function corresponds to the trailing edges in reflection spectra.

### 3.4. *Thermodynamics properties*

We have investigated the temperature-dependent of some physical properties of $Cu_3TMS_4$ (*TM* =V, Nb, Ta) by using quasi-harmonic Debye approximation [28] which is a standard method for calculating the thermodynamics properties. It is shown by several authors that their calculated data using this method are in good agreement with experimental values [9,29]. Moreover our calculated Debye Temperatures using this approximation are in good agreement with the values calculated from elastic modulus as discussed later. The volume and total energy of the unit cell of the studied materials calculated using CASTEP code based on DFT method within the generalized gradient approximation, were used as input data in Gibbs program. Here we computed the bulk modulus, Debye temperature and specific heats at different temperatures and pressures for the first time and there is no noticeable difference for the temperature and pressure dependence properties among the herein studied materials.

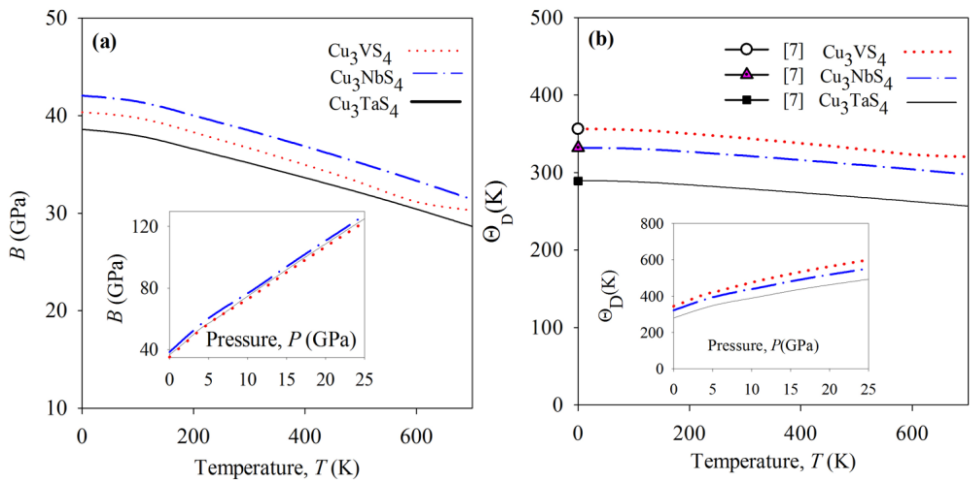

Fig. 5. Temperature dependence of bulk modulus, *B* (a) and Debye temperature, $\Theta_D$ (b) of $Cu_3TMS_4$ (*TM* =V, Nb, Ta). *Insets* show pressure variation. Zero temperature $\Theta_D$ derived from elastic data[7] are shown for comparison.

The temperature variation of isothermal bulk modulus of $Cu_3TMS_4$ ($TM$ =V, Nb, Ta) at $P = 0$ GPa are presented in Fig. 5(a) and the *inset* of which shows $B$ as a function of pressure at room temperature. It is seen that the bulk modulus is nearly constant from 0 to 100 K. For $T > 100$ K, the bulk modulus decreases with the increasing of temperature at a given pressure and increases with pressure at a given temperature and the shape of the curve is nearly linear. This means that the compressibility increases with increasing temperature at a given pressure and decreases with pressure at a given temperature. The results are due to the fact the effect of increasing pressure on material is similar as decreasing temperature of material, which means that the increase of temperature of the material causes a reduction of its hardness.

Fig. 5(b) displays the temperature dependence of Debye temperature $\Theta_D$ of $Cu_3TMS_4$ ($TM$ =V, Nb, Ta) at $P = 0$ GPa along with the data obtained by Espinosa-García *et al.* [7]. The *inset* of the figure shows $\Theta_D$ as a function of pressure at room temperature. At fixed pressure, Debye temperature decreases with increasing temperatures and at fixed temperature $\Theta_D$ increases with increasing pressure, indicating the change of the vibration frequency of particles under pressure and temperature effects. Most other solids have weaker bonds and far lower Debye temperatures; consequently, their heat capacities have almost reached the classical Dulong–Petit value of $3R$ at room temperature as can be seen from Fig. 6(a). If it seems that the harder is the solid, the higher is the Debye temperature, and the slower is the solid to reach its classical $C_V$ of $3R$, this is not a coincidence. One observes that the value of $\Theta_D$ smaller for $Cu_3TaS_4$ and larger for $Cu_3VS_4$. Our calculated values of $\Theta_D$ at $T = 0$ are 356, 332, and 289 for $Cu_3TMS_4$ ($TM$ =V, Nb, Ta), respectively which are in good agreement with the result obtained by Espinosa-García *et al.* [7] derived from elastic moduli.

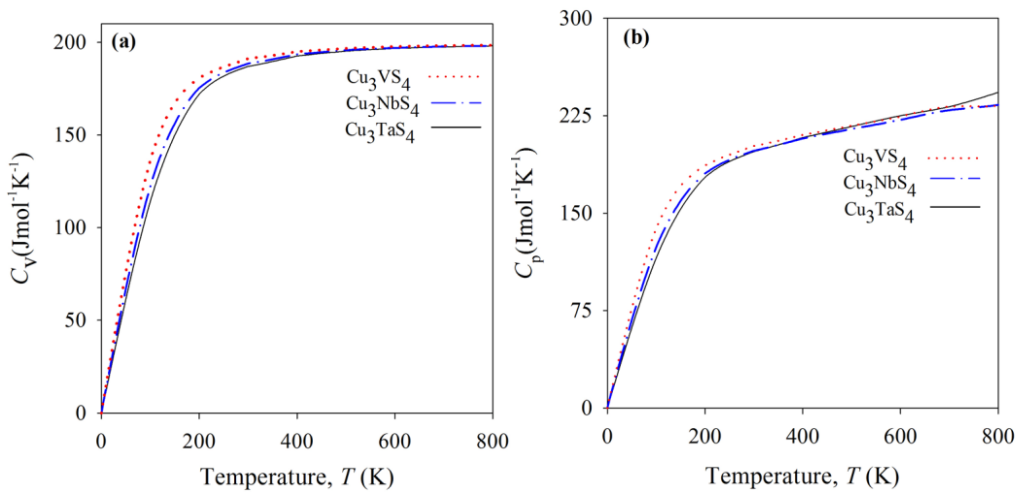

Fig. 6. Temperature dependence of lattice specific heat at constant volume (a) and specific heat at constant pressure (b) of $Cu_3TMS_4$ ($TM$ =V, Nb, Ta).

The lattice heat capacity of a substance is a measure of how well the substance stores heat. The temperature dependence of $C_V$ is governed by the details of vibrations of the atoms and could be determined from experiments. It is worthwhile to outline that the Debye model correctly predicts the low-temperature dependence of the heat capacity at constant volume, which is proportional to $T^3$ [30]. It also recovers the Dulong–Petit law at high temperatures [31]. The heat capacities at constant-volume ($C_V$) and constant-pressure ($C_P$) of $Cu_3TMS_4$ (*TM* =V, Nb, Ta) as a function of temperature are displayed in Fig. 6 (a, b). The temperature is limited to 800 K to reduce the possible effect of anharmonicity. The heat capacity $C_V$ and $C_P$ both increase with the applied temperature, and there is no noticeable difference among the materials studied here. As temperature increases phonon thermal softening occurs and as a result the heat capacities increase with increasing temperature. The difference between $C_P$ and $C_V$ for the phase is given by $C_P - C_V = \alpha_V^2(T) BTV$ ($\alpha_V$ = volume thermal expansion coefficient), which is due to the thermal expansion caused by anharmonicity effects. In addition, due to the anharmonic approximations of the Debye model used here, for higher temperatures, the anharmonic effect on $C_V$ is suppressed, and $C_V$ is very close to the Dulong–Petit limit which is common to all solids at high temperature. Above 300 K, $C_V$ increases slowly with temperature and gradually approaches the Dulong–Petit limit ($C_V = 3nNk_B$ = 199.5 J/mol-K). These results reveal that the interactions between ions in the sulvanites have great effect on heat capacities especially at low *T*.

## 4. Conclusions

A first time investigations of the optical and thermodynamics properties including a revisit of the structural, elastic, electronic properties at zero temperature and pressure of three sulvanites $Cu_3TMS_4$ (*TM* =V, Nb, Ta) have been carried out and the results compared with existing data where available. The analysis of elastic constants reveals that the compounds under consideration are mechanically stable. The calculated band structures show the semiconducting behavior of these materials with comparable band gaps from other study. The dielectric function, refractive index, absorption spectrum, conductivity, reflectivity and energy-loss spectrum have been calculated and discussed in detail. The static dielectric constants have the values 8.56, 7.32, and 7.10 for $Cu_3VS_4$, $Cu_3NbS_4$, and $Cu_3TaS_4$, respectively indicating that they are dielectric materials. The absorption coefficient and photoconductivity curves show the semiconducting behavior. The reflectivity spectra show that the compounds are mostly transparent in the visible range. For $Cu_3TaS_4$ the calculated refractive index and reflectivity coefficient both roughly follow the available low energy measured data. Finally, the first-time results on temperature and pressure dependence bulk modulus, Debye temperature and heat capacities are obtained and discussed. The variation of Debye temperature with temperature and pressure reveals that the vibration frequencies of the particles in these materials are changeable. The heat capacities increase with increasing temperature, which shows that phonon thermal softening occurs when the temperature increases.